\documentclass[aps,pra,showpacs,preprintnumbers,amsmath,amssymb,reprint,superscriptaddress]{revtex4-2}
\usepackage{graphicx,dcolumn,bm}
\usepackage[utf8]{inputenc}
\usepackage[T1]{fontenc}
\usepackage{mathptmx}
\usepackage{bm}
\usepackage{siunitx}
\usepackage{booktabs}
\usepackage{dcolumn}
\usepackage{xcolor}
\newcolumntype{d}[1]{D{.}{.}{#1}}
\begin{document}

\preprint{}

	\title[On the $\Lambda$-doubling in the B$^3\Pi_1$ state of TlF]{On the $\Lambda$-doubling in the B$^3\Pi_1$ state of TlF}
	
    \author{Gerard Meijer}
	\email{meijer@fhi-berlin.mpg.de} 
	\affiliation{Fritz-Haber-Institut der Max-Planck-Gesellschaft, Faradayweg 4-6, 14195 Berlin, Germany}
	\author{Boris G. Sartakov}
	\affiliation{Prokhorov General Physics Institute, Russian Academy of Sciences, Vavilovstreet 38, 119991 Moscow, Russia}
	
\date{\today}
\begin{abstract}
	Thallium monofluoride (TlF) is a prime candidate molecule for precision measurements aimed at discovering new physics. Optical cycling on the B $\leftarrow$ X transition around 271 nm enhances this potential. Hyperfine resolved ultraviolet spectra have been reported to determine the degree of rotational level mixing in the B-state and the efficiency of laser cooling on the B $\leftarrow$ X transition \cite{Norrgard2017}. Using these high-resolution spectra, the hyperfine structure in the B-state of TlF is re-analysed and the magnitude of the $\Lambda$-doubling in the B-state is discussed.
\end{abstract}
	
\maketitle

\section{\label{sec:introduction} Introduction and motivation}

The ultraviolet spectrum of thallium fluoride was recorded and comprehensively analysed for the first time by Howell in 1937 \cite{Howell1937}. He identified the band system centered around 271 nm as a $^3\Pi_0$ -- X$^1\Sigma^+$ system. Some 1700 cm$^{-1}$  below the $^3\Pi_0$ state he identified a $^3\Pi_1$ state. He rationalized why these states should appear in this energetic order (inverted) and that these are the only two low-lying states that can be reached from the electronic ground state \cite{Howell1937}.

The first rotational analyses of the bands of the $^3\Pi_0$ -- X$^1\Sigma^+$ and $^3\Pi_1$ -- X$^1\Sigma^+$ systems were performed by Barrow and co-workers in 1958 \cite{Barrow1958}. In this work the assignment of the two electronically excited states was interchanged: the lowest one was referred to as the A$^3\Pi_0$ state and the upper one as the B$^3\Pi_1$ state. In their paper, this interchanged assignment is not commented upon. They do remark, however, that the A-X system consists of $R$ and $P$ branches ``{\it spreading in opposite directions from well marked origin gaps}'', thereby implicitly stating that there is no $Q$-branch in the A-X system. They also make a one-sentence remark on the observation of a clear $Q$-branch in the B-X system \cite{Barrow1958}. Interestingly, already in 1950, Herzberg writes about the A- and B-state of TlF in a footnote to the 80-page Table of Molecular Constants in his classic book that ``{\it Howell exchanges $^3\Pi_1$ and $^3\Pi_0$ but the B $\leftarrow$ X system has strong $Q$-branches and must therefore have $\Delta \Omega$= $\pm$1}'' \cite{Herzberg1950}.

Since then, Tiemann and coworkers have analysed the rotational structure in the B $\leftarrow$ X system up to high rotational quantum numbers (above $J=100$) and they have studied the predissociation of the B-state \cite{Wolf1987}. In their analysis, they extracted for the first time values for the $\Lambda$-doubling in the B-state and reported effective Dunham parameters for this \cite{Tiemann1988}.

Our interest in the detailed energy level structure in the B-state of TlF was triggered by the observed anomolously large hyperfine splitting of the $J=1$ level \cite{Norrgard2017}. This splitting brings the lowest lying hyperfine levels in the B-state close in energy to where a fictitious $J=0$ level would be expected. We wondered, therefore, whether the B-state of TlF should be described as a $^3\Pi_0$ state after all. Our interest was also triggered by our recent spectroscopic study on the related AlF molecule. In that study, we measured and modelled the hyperfine splittings and $\Lambda$-doublings in all three $\Omega$-manifolds of the electronically excited a$^3\Pi$ state to kHz precision, using a standard Hamiltonian \cite{Truppe2019}. The model used in ref. \cite{Norrgard2017} to describe the TlF spectra is slightly different and although the match obtained between the observed and simulated low-$J$ Q-branch spectra of both TlF isotopes is good, it is not as perfect as one would expect it to be. We therefore set out to apply the same formalism as used for AlF to the published spectroscopic data on the TlF molecule \cite{Norrgard2017}. As described here, the hyperfine structure in the B-state unambiguously confirms that this state cannot be a $^3\Pi_0$ state. An excellent match between the observed and calculated hyperfine resolved spectra is found when the standard Hamiltonian for an isolated $^3\Pi_1$ state is used. Accurate information on the $\Lambda$-doubling as well as on the $g_F$-factors of the low-$J$ levels in the B-state of TlF is obtained.

\section{\label{sec:Hyperfine Structure} Fit of the Hyperfine Structure}

For our analysis, we used all the measured values for the hyperfine splittings of the $^{205}$TlF isotope as reported in Table I and Table II of ref. \cite{Norrgard2017}, 48 for $e$-levels and 83 for $f$-levels. Different from the procedure followed in ref. \cite{Norrgard2017}, we included the $e$- and $f$-levels in a single fit \cite{Truppe2019}. We did not include the reported hyperfine splittings of the less abundant $^{203}$TlF isotope, as fewer of these are tabulated and as there is very limited information on the $f$-levels for low values of $J$, which could bias the overall fit. Instead, after completing the fitting procedure, we used the isotopic scaling rules of the parameters found for $^{205}$TlF to determine the parameters for $^{203}$TlF, and to simulate the Q-branch spectra of both isotopes.

The hyperfine Hamiltonian, $H_{HFS}$, for each of the nuclei with nuclear spin ${\bf{I}}$ ($I_{Tl}$ = $I_F$ = $1/2$) of TlF can be written as
\begin{equation}
H_{HFS}=a\,L_z\,I_z+b_F\,{\bf{I}}\cdot {\bf{S}}+\frac{1}{3}\,c\,(3\,S_z\,\,I_z-{\bf{S}}\cdot{\bf{I}}),
\end{equation}
with the orbital, $a$, the Fermi contact, $b_F$, and the dipolar, $c$, parameters as originally defined by Frosch and Foley \cite{Frosch1952,Carrington2003}. For a given electronic state, with a component $L_z=\Lambda$ of the total electron orbital angular momentum along the internuclear axis, with a total electron spin $S$ and with quantum number $\Omega$ -- where $\Omega$ = $\Lambda$ + $\Sigma$, with $\Sigma$ the projection of $\bf{S}$ along the internuclear axis -- the Hamiltonian can be expressed as
\begin{equation}
H_{HFS}=I_z\left[a\,\Lambda+b_F(\Omega-\Lambda)+\frac{2}{3}\,c\,(\Omega-\Lambda)\right].
\end{equation}
In the B-state of TlF, we can be certain that $\Omega=1$ ($\it{vide~infra}$) but the quantum numbers $\Lambda$ and $S$ are not well determined. It is commonly assumed that for the B-state $\Lambda=1$ and $S=1$ but any electronic state with $\Lambda=1, 2, \dots$ and with the appropriate value for the electron spin has an $\Omega=1$ component that can be admixed. The relatively short lifetime of the B-state of 99 $\pm$9~ns \cite{Hunter2012} suggests that $S=0$ electronic states contribute significantly to the B-state wavefunction. We therefore follow here the notation that is also used in ref. \cite{Norrgard2017}, and write
\begin{equation}
H_{HFS}=h_{\Omega}\,I_z,
\end{equation}
implying that $h_1$ $\equiv$ $a$ when $\Lambda$=1 whereas $h_1$ will have contributions with different weights from the $a$, $b_F$ and $c$ terms when $\Lambda$ and $S$ have different values.

The lowest two, isolated $F=0$ and $F=1$ hyperfine levels in the B-state of TlF are located at an energy close to where the $J=0$ level would be expected if the B-state were a $^3\Pi_0$ state. We considered it important to use the analysis of the observed hyperfine structure to check whether the B-state might be a $\Omega=0$ state after all. For this, we reassigned the observed rotational transitions \cite{Norrgard2017} to those of a $^3\Pi_0$ $\leftarrow$ $^1\Sigma^+$ band and analysed the resulting hyperfine splittings with the Hamiltonian for an isolated $^3\Pi_0$ state. No satisfactory agreement could be obtained when the main two parameters, $h_0(Tl)$ and $h_0(F)$, were used. In an effort to better match the resulting hyperfine splittings, terms describing the interaction of each nucleus with the pure rotational angular momentum (parameters $C_I(Tl)$ and $C_I(F)$) as well as nuclear spin-spin interaction terms (a scalar one and a tensorial one, described with the parameters $D_0$ and $D_1$, respectively) were included \cite{Truppe2019}. Also in this case, no satisfactory agreement between theory and experiment could be obtained. We conclude, therefore, that based on the observed hyperfine structure, the B-state of TlF cannot be described as a $^3\Pi_0$ state.

We then analysed the hyperfine structure in the B-state of TlF using a standard Hamiltonian for an isolated $^3\Pi_1$ state \cite{Truppe2019}. This Hamiltonian is identical to the one for a $^1\Pi$ state. As both nuclei have a nuclear spin of $1/2$, one can formally not distinguish which of the two main $h_1$ parameters belongs to which nucleus; it is expected, however, that $h_1(Tl)$ is (much) larger than $h_1(F)$ and they are assigned accordingly. In our Hamiltonian, we also include the effective nuclear spin-rotation parameter $C_I$ as well as the corresponding $\Lambda$-doubling contribution $C'_I$ as defined by Brown and coworkers \cite{Brown1978}. Considering the relative magnitude of $h_1(Tl)$ and $h_1(F)$, these higher order $C_I$ and $C'_I$ terms are only included for the Tl nucleus. The $C'_I(Tl)$ term couples $e$- and $f$-levels and therefore, both sets of levels need to be treated in a single fit. It is this term that causes the so-called $"b"$-splitting \cite{Norrgard2017} to be different for $e$- and $f$-levels and as these splittings are tabulated up to high $J$-values, the value of $C'_I(Tl)$ can be accurately determined. 

\begin{table}[hbt!]
	\centering
	\begin{tabular}{|c|c|c|c|} \hline
		Parameter & value (MHz) & SD & SD$\cdot$ sqrt(Q)\\
		\hline	
		$B              $ &      6687.879 &          --  &        --  \\
		$q              $ &         2.423 &          -- &         --  \\
		$D              $ &      0.010869 &  --   &  --  \\
		$ H             $ &      8.1 $\cdot 10^{-8}$  & --   &  --  \\
		$h_1(Tl)      $ &        28789  &      34  &      42   \\
	    $h_1(F)       $ &             861 &         17   &      20   \\	
		$C_I(Tl)       $ &         -7.83  &         0.43  &         1.57   \\
		$C'_I(Tl)       $ &         11.17  &          0.85  &         3.00 \\
		$\Delta \nu   $ &        2571 &          --  &        -- \\
		\hline
	\end{tabular}
	\caption{\label{tab:HFS-Constants} Rotational constant $B$, $\Lambda$-doubling parameter $q$, rotational centrifugal distortion parameters $D$ and $H$ and hyperfine parameters $h_1(Tl)$, $h_1(F)$, $C_I(Tl)$ and $C'_I(Tl)$ for $^{205}$TlF, as obtained from the best fit to the 131 hyperfine splittings of $^{205}$TlF listed in Table I and Table II of ref. \cite{Norrgard2017}. The four hyperfine parameters are given together with their standard deviation (SD) and the product of SD and $\sqrt{Q}$ (all values in MHz). The other parameters are kept fixed in the fit. The standard deviation of the fit is 8 MHz. The parameter $\Delta \nu$ is the shift of the vibrational band origin of $^{203}$TlF relative to $^{205}$TlF.}
\end{table}

The parameters resulting from a best fit to the hyperfine splittings of $^{205}$TlF are presented in Table~\ref{tab:HFS-Constants}. For the four hyperfine parameters, the standard deviation (SD) as well as the product of the standard deviation with the square-root of the quality-factor Q is given; the latter is the better measure for the accuracy with which each parameter is determined in a fitting procedure in which various parameters are correlated \cite{Watson1977}. The standard deviation of the fit is 8~MHz. 

The hyperfine parameters $h_1(Tl)$ and $h_1(F)$ are within the error bars the same as those found in the study by Norrgard $\it{et~al.}$ \cite{Norrgard2017}. In that study the hyperfine parameters were fitted separately for $e$- and $f$-levels, i.e. two more free parameters were used. A comparison of the higher order hyperfine parameters is not straightforward, as in ref. \cite{Norrgard2017} the $C'_I(Tl)$ term was not included, but two separate $C_I(Tl)$ terms for the $e$- and $f$-levels were used instead; the value we find for $C_I(Tl)$ is the average of the two separate values reported there \cite{Norrgard2017}. As mentioned before, the seperate fitting for the $e$- and $f$-levels is formally not valid and can result in a wrong correlation between hyperfine constants, leading to unrealistic error bars.

The rotational constants $B$, $D$ and $H$ as well as the $\Lambda$-doubling parameter $q$ for $^{205}$TlF are kept fixed in the fit of the hyperfine splittings at the tabulated values. The $\Lambda$-doubling is generally described by three terms in the Hamiltonian, parametrized by the $o$-, $p$- and $q$-parameters \cite{Brown1979}. In the case of an isolated $\Omega=1$ state only the $q$-term has non-vanishing matrix elements between sub-levels of the same $J$. As this term yields an energy difference between $e$- and $f$-levels given by $q J (J+1)$ its effect can be incorporated by using different rotational constants for the $e$- and $f$-levels. This is the approach used in ref. \cite{Norrgard2017} and the value for $q$ that is given in Table~\ref{tab:HFS-Constants} is the difference of their reported $B$-values for the $e$- and $f$-levels; the value for $B$ that is given in Table~\ref{tab:HFS-Constants} is the average of their $B$-values. The centrifugal distortion parameters $D$ and $H$ are those reported for the $f$-levels in ref. \cite{Norrgard2017}. 

We have used these $B$, $D$, $H$ and $q$ parameters to simulate the Doppler limited absorption spectra of the $v'=0$ $\leftarrow$ $v''=0$ band recorded by Tiemann and coworkers, tabulated in the PhD thesis of Wolf \cite{WolfPhD1987}. There, the frequencies of many lines in the $R$- and $P$-branch are listed, but in the $Q$-branch only the frequencies for isolated lines in a limited interval of high $J$-values are given. Our simulations show that their labeling of the $Q$-lines is one quantum number off. As a consequence, the value for the $\Lambda$-doubling parameter that they extracted for the $v''=0$ level of $q$ = 0.1 $\pm$ 0.3~MHz, is incorrect \cite{Tiemann1988}.

\begin{figure}[hbt!]
	\centering
	\includegraphics[width=1.0\linewidth]{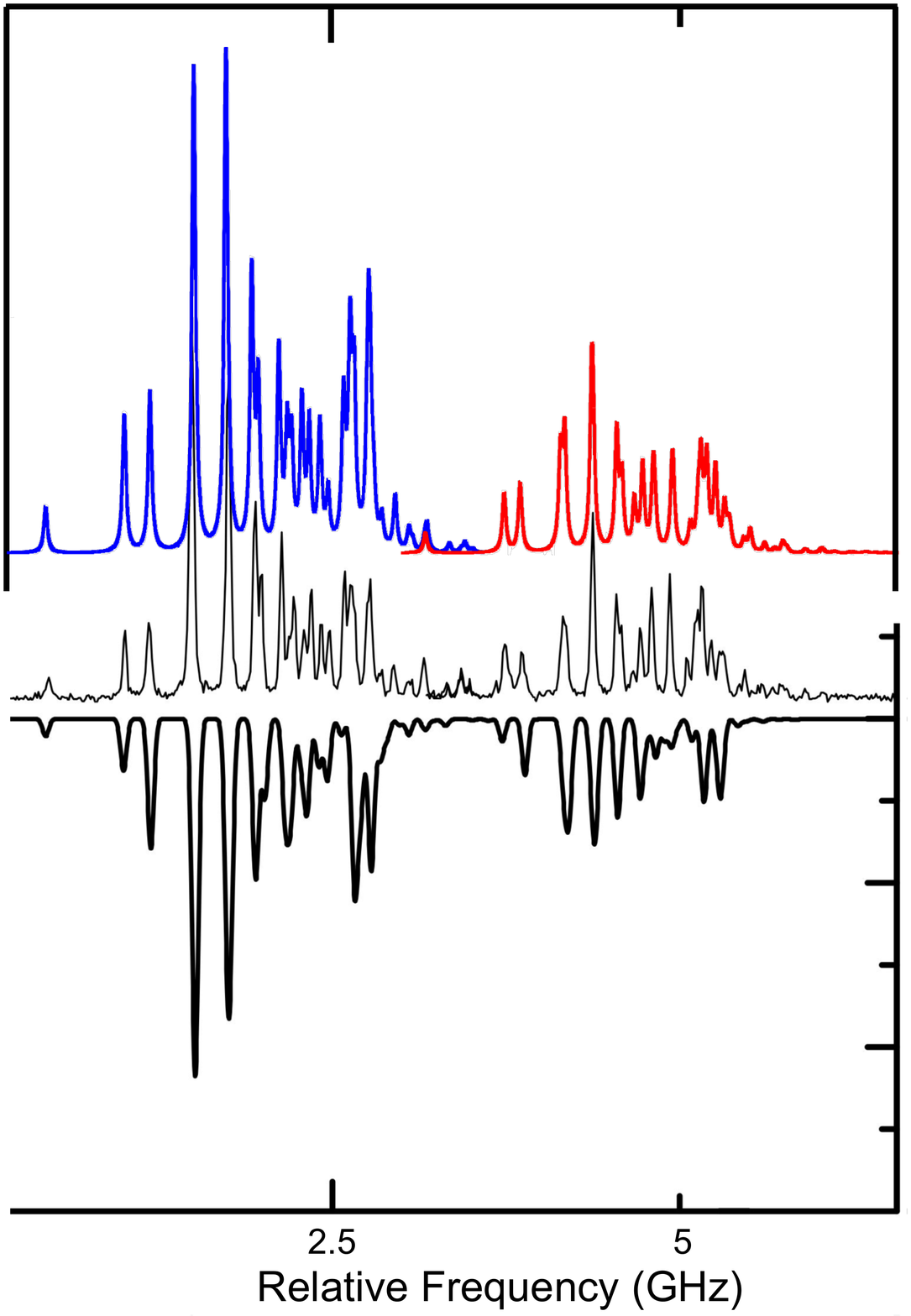}
	\caption{Reproduction of the measured $Q$-branch spectra of the B$^3\Pi_1$, $v$=0 $\leftarrow$ X$^1\Sigma^+$, $v''$=0 band of the $^{205}$TlF and $^{203}$TlF isotopes (middle) together with the simulation (lower), as presented in Fig. 2(b) of ref. \cite{Norrgard2017}. The upper spectrum shows our simulated $Q$-branch spectrum for $^{205}$TlF (blue) and $^{203}$TlF (red), using the parameters listed in Table~\ref{tab:HFS-Constants}. In our simulation, a rotational temperature of 7~K and a Lorentzian linewidth (FWHM) of 30~MHz is taken. The parameters for the $^{203}$TlF isotope are obtained from the parameters in Table I by using the isotopic scaling rules.}
	\label{fig:Q-Branches}
\end{figure}

The hyperfine resolved fluorescence excitation spectrum of the $Q$-branch of the B$^3\Pi_1$, $v=0$ $\leftarrow$ X$^1\Sigma^+$, $v''=0$ band of both the $^{205}$TlF and $^{203}$TlF isotope is presented in Fig. 2(b) of ref. \cite{Norrgard2017}, together with their simulated spectrum. Using the parameters listed in Table~\ref{tab:HFS-Constants}, assuming a rotational temperature of 7~K and a Lorentzian lineshape with a full width at half maximum of 30~MHz, the simulated spectrum shown in Figure~\ref{fig:Q-Branches} (above part of their reproduced Fig. 2(b)) is obtained. For this simulation, the rotational constant of $^{205}$TlF in the X$^1\Sigma^+$, $v''=0$ state is taken from microwave spectroscopy as 6667.4~MHz  \cite{Barrett1958,Tiemann1988}. The hyperfine splittings in the electronic ground-state are below 1~MHz and the hyperfine levels belonging to a certain $J''$ are taken to be degenerate. The blue part in the simulated spectrum is the contribution from $^{205}$TlF and the red part is from $^{203}$TlF; both isotopes are assumed to be present in their natural abundance. The shift $\Delta \nu = 2571$~MHz of the vibrational band origin of $^{203}$TlF relative to $^{205}$TlF is determined from the isotope-splittings tabulated in ref. \cite{Norrgard2017}.
The simulated spectrum reproduces even the finest details of the experimental spectrum, for both isotopes. This not only attests to the quality of the experimental spectrum, but it also unambiguously shows that the B-state of TlF is an $\Omega =1$ state.

\section{\label{sec:Lambda Doubling} $\Lambda$--doubling for $J=1,2$}

In an isolated $\Omega=1$ state, the magnitude of the $\Lambda$-doubling for a given $J$-level is normally described by the term $q J(J+1)$, as stated before. In the B-state of TlF, however, the contribution of the $C'_I$ terms dominates the splitting between the opposite parity components of a given $F$-level for low $J$-values. For the $J=1$, $F=0$ level the analytical expression for the $\Lambda$-doubling is 
\begin{equation}
E_e - E_f = 2q + C'_I(Tl) + C'_I(F).
\end{equation}
In fitting the hyperfine splittings, we did not include the $C'_I(F)$ term, i.e. we kept the value of $C'_I(F)$ fixed to zero. If we include $C'_I(F)$ in the fit, we find a value of 0.01 $\pm$ 0.30~MHz. Including this term, therefore, does not change the value of the $\Lambda$-doubling, but only adds a small uncertainty. The value that we find for the $\Lambda$-doubling of the $J=1$, $F=0$ level is 16 $\pm$ 1~MHz. Most importantly, the magnitude of the $\Lambda$-doubling does not depend on the (error bar of) the large $h_1(Tl)$ and $h_1(F)$ parameters.

\begin{table}[hbt!]
	\centering
\begin{tabular}{|c|c|c|c||c|c|c|c|}
  \hline
  $(J,F_1,F)$ & $\Delta E_{\Lambda}$ (MHz) & SD &$g_F$& $(J,F_1,F)$ & $\Delta E_{\Lambda}$ (MHz) & SD&$g_F$ \\
  \hline
  (1,1/2,0) &  16 & 1  & 0    & (2,5/2,2) & -14 & 2 & 0.11 \\
  (1,1/2,1) &  16 & 1  & 0.32 & (2,5/2,3) & -14 & 2 & 0.08 \\
  (1,3/2,1) & -18 & 1  & 0.25 & (2,3/2,1) &  15 & 4 & 0.43 \\
  (1,3/2,2) & -18 & 1  & 0.14 & (2,3/2,2) &  14 & 4 & 0.25 \\
  \hline
\end{tabular}
\caption{\label{tab:Lambda-Doubling} Energy difference $\Delta E_{\Lambda}$ = $E_e$ -- $E_f$ between the $e$- and $f$-components of the four hyperfine levels of the two lowest rotational levels in the B-state, labeled by ($J$, $F_1$, $F$), following the nomenclature in ref. \cite{Norrgard2017}. The $\Lambda$-doublet splittings are given together with their standard deviation (SD) (all values in MHz). The calculated $g_F$-values, assuming an isolated $^3\Pi_1$ state or a $^1\Pi$ state, are given in a separate column.}
\end{table}

For the $F\neq 0$ levels, the analytical expressions for the $\Lambda$-doubling are more involved, and only the numerical values resulting from the fit are given in Table~\ref{tab:Lambda-Doubling} for the eight lowest energy $F$ levels. From this Table, it is seen that the values for the $\Lambda$-doubling are positive for the $F_1$ = $J - 1/2$ levels and negative for the $F_1$ = $J + 1/2$ levels. Interestingly, the magnitude of the $\Lambda$-doubling is seen to drop slightly in going from $J=1$ to $J=2$. The separation between the opposite parity components of these $F$-levels is only about ten times their homogeneous linewidth \cite{Hunter2012}. The $F\neq 0$ levels will experience a first-order Stark-shift in an external electric field. Assuming that the electric dipole moment in the B-state has a value that is comparable to the 4.2~Debye in the X$^1\Sigma^+$ state \cite{Boeckh1964}, electric fields of a few V/cm will already lead to significant mixing of the opposite parity components of these $F$ levels.

For the shifting and splitting of the $F$-levels in a magnetic field, the magnetic $g_F$-factors need to be known. We know that in the B-state $\Omega = 1$ and we have calculated the $g_F$-factors using the formalism for an isolated $^3\Pi_1$ state as well as for a $^1\Pi$ state. We find the same values for these two cases, confirming that these two models are equivalent, and these $g_F$-values are given in Table~\ref{tab:Lambda-Doubling}. It should be noted that the presence of other electronic states close to the $^3\Pi_1$ (or $^1\Pi$) state can potentially influence these $g_F$-values.

\section{\label{sec:Conclusions} Concluding remarks}

The rigorous approach to incorporate the $\Lambda$-doubling and higher-order hyperfine structure terms in the Hamiltonian, in a combined fit of the $e$- and $f$-levels, is seen to describe the energy level structure in the B-state of TlF very well, unambiguously demonstrating that this is an $\Omega=1$ state. It is remarkable that this state can be so well described by the Hamiltonian of an isolated $^3\Pi_1$ state, even up to rotational energies of 0.25~eV above the lowest level, given the high density of electronically excited states nearby \cite{Balasubramanian1985,Zou2008}. The large value for $h_1$ would be extraordinary for light diatomics, but might be common for molecules with heavy nuclei. Relativistic effects can lead to a decrease of the radius of the electron orbit and might cause this large value of $h_1(Tl)$. As a result, the values for $C_I(Tl)$ and $C'_I(Tl)$ are also larger than for light diatomics; the value for $C'_I(Tl)$ found here is about three orders of magnitude larger than the value for $C'_I(Al)$ found in the a$^3\Pi$ state of AlF \cite{Truppe2019}. The overall picture of the hyperfine structure in the B-state is the same as reported in ref. \cite{Norrgard2017} and, in particular, the rovibrational branching ratios reported there are not found to be different in our more rigorous analysis. We do find, however, that the experimental data also contain accurate information on the separation and ordering of the opposite parity components of the lowest $F$-levels, i.e. the levels that are most important for the laser cooling experiments \cite{Hunter2012}.

\section{\label{sec:Acknowledgment} Acknowledgment}

We acknowledge the constructive scientific discussion with Dave DeMille, Eric Norrgard and Larry Hunter on the details and differences of their and our analysis of the excellent experimental data that they presented in ref. \cite{Norrgard2017}.

\bibliographystyle{aipnum4-1}

\begin{thebibliography}{}
	
	\bibitem{Norrgard2017}
	E. B. Norrgard, E. R. Edwards, D. J. McCarron, M. H. Steinecker, D. DeMille, S. S. Alam, S. K. Peck, N. S. Wadia, and L. R. Hunter,
	Physical Review A {\bf 95}, 062506 (2017). 
	
	\bibitem{Howell1937}
	H. G. Howell, Proceedings of the Royal Society of London A {\bf 160}, 242 (1937).
	
	\bibitem{Barrow1958}
	R. F. Barrow, H. F. K. Cheall, P. M. Thomas, and P. B. Zeeman, Proceedings of the Physical Society {\bf 71}, 128 (1958).
	
	\bibitem{Herzberg1950}
	G. Herzberg, {\it Molecular Spectra and Molecular Structure. I. Spectra of diatomic molecules.}, 2nd ed. (Van Nostrand Reinhold Company Inc., 1950).
	
	\bibitem{Wolf1987}
	U. Wolf and E. Tiemann, Chemical Physics Letters {\bf 133}, 116 (1987).
	
	\bibitem{Tiemann1988}
	E. Tiemann, Molecular Physics {\bf 65}, 359 (1988).
	
	\bibitem{Truppe2019}
	S. Truppe, S. Marx, S. Kray, M. Doppelbauer, S. Hofs\"ass, H. C. Schewe, N. Walter, J. P\'erez-R\'ios, B. G. Sartakov, and G. Meijer,
	Physical Review A {\bf xx}, xx (2019): arXiv:1908:11774.
	
	\bibitem{Frosch1952}
	R. A. Frosch and H. M. Foley, Physical Review {\bf 88}, 1337 (1952).
	
	\bibitem{Carrington2003}
	A. Carrington and J. M. Brown, {\it Rotational spectroscopy of diatomic molecules} (Cambridge University Press, 2003).

	\bibitem{Hunter2012}
	L. R. Hunter, S. K. Peck, A. S. Greenspon, S. S. Alam, and D. DeMille, Physical Review A {\bf 85}, 012511 (2012).
	
	\bibitem{Brown1978}
	J. M. Brown, M.  Kaise, C. M. L. Kerr, and D. J. Milton, Molecular Physics {\bf 36}, 553 (1978).

	\bibitem{Watson1977}
	J. K. G. Watson, Journal of Molecular Spectroscopy {\bf 66}, 500 (1977).
	
	\bibitem{Brown1979}
	J. M. Brown and A. J. Merer, Journal of Molecular Spectroscopy {\bf 74}, 488 (1979).

	\bibitem{WolfPhD1987}
	U. Wolf, {\it Laserspektroskopische Untersuchungen zur Pr\"adissoziation der IIIa-VII-Verbindungen durch Tunneleffekt}, Ph.D. thesis, Universit\"at Hannover (1987).
	
	\bibitem{Barrett1958}
	A. H. Barrett and M. Mandel, Physical Review {\bf 109}, 1572 (1958).

	\bibitem{Boeckh1964}
	R. von Boeckh, G. Gr\"aff, and R. Ley, Zeitschrift f\"ur Physik {\bf 179}, 285 (1964). 
	
	\bibitem{Balasubramanian1985}
	K. Balasubramanian, The Journal of Chemical Physics {\bf 82}, 3741 (1985). 
	
	\bibitem{Zou2008}
	W. Zou and W. Liu, Journal of Computational Chemistry {\bf 30}, 524 (2008). 

\end{thebibliography}

\end{document}